\documentclass[12pt,preprint]{aastex}

\usepackage{graphicx}

\newcommand{\bolder}[1]{{\mbox{\boldmath${#1}$\unboldmath}}}
\newcommand{\uvec}[1]{{\bolder{\hat{#1}}}}

\shortauthors{Bailin \& Steinmetz}
\shorttitle{Tidal Torques and Galactic Warps}

\begin{document}

\title{Tidal Torques and Galactic Warps}

\author{Jeremy Bailin\altaffilmark{1,2,3} and Matthias Steinmetz\altaffilmark{1,2,4,5}}

\altaffiltext{1}{Steward Observatory, University of Arizona,
	933 North Cherry Ave, Tucson, AZ 85721, USA}
\altaffiltext{2}{Astrophysikalisches Institut Potsdam, An der Sternwarte 16,
	14482 Potsdam, Germany}
\altaffiltext{3}{jbailin@as.arizona.edu}
\altaffiltext{4}{msteinmetz@aip.de}
\altaffiltext{5}{Alfred P. Sloan Fellow and David and Lucile Packard Fellow}

\begin{abstract}
We investigate how galactic disks react to external tidal torques.  We
calculate the strength and radial dependence of torques on disks that
arise from a misalignment between the disk and the main axis system of
a flattened dark matter halo. Density profile, misalignment and
flattening of the halo are chosen to match the corresponding values
typically found for dark matter halos in large cosmological N-body
simulations. We find that except for in the very inner regions, the
torques are well-described by a power law of the form $\tau \propto
r^{-2.5}$. For torques as they arise in typical cosmological
settings, the magnitude of the torque is large enough for the entire disk
to react to the torque in less than the Hubble time.  We demonstrate
analytically that disks which are originally located in the $xy$-plane and which are
subjected to a torque around the $x$-axis tilt around the $y$-axis, as
also found in fully non-linear N-body simulations.
We further demonstrate that that the torque causes the radius of a
chosen particle to increase with time.
 Investigations of tilting disks which treat the disk
as a set of solid rings thus may systematically overestimate the effects of the
torque by a factor of two.  For torques of the form we investigate,
the inner regions of the disk react to the torque faster than the
outer regions, resulting in a trailing warp. We then study the effect
of the self-gravity of the disk in such a scenario using numerical
N-body models.  Self-gravity flattens out the inner regions of the
disk, but these regions are tilted with respect to their initial
plane followed by a non-flat outer region whose tilt decreases with radius. 
The ``warp radius,'' which marks the end of the inner flat
disk, grows throughout the disk at a rate that depends only on the
strength of the torque and the local surface density of the disk.
\end{abstract}


\section{Introduction}

Many edge-on disk galaxies show integral-sign or S-shaped warps, where
the majority of the disk is planar but where the outer region of the
disk lies above that plane on one side of the galaxy and below the
plane on the other \citep{binney92}. The Milky Way is warped both
in neutral hydrogen \citep[e.g.][]{diplas and savage91} and in the
stellar distribution \citep{reed96,lopez-corredoira et al02b}. \citet{reshetnikov and combes98}
estimate that half of all disk galaxies have optical warps, and most
HI disks which extend beyond the optical disks appear to be warped
\citep{bosma81,briggs90,christodoulou et al93}.

A number of methods have been proposed for creating and maintaining warps.
Several authors \citep{lynden-bell65,sparke and casertano88} have
suggested that the system of discrete particles which make up the
disk may have normal bending modes that could be excited. \citet{battaner et al}
have investigated magnetic fields as a cause of warps, while \citet{toomre83}
and \citet{dekel and shlosman83}  suggested that disks askew in flattened
dark matter halos could develop long-lived warps, assuming the radial
profile of the halo were appropriately fine-tuned.

Following on the idea that infalling material will shift the angular
momentum of a galaxy \citep{ryden88,quinn and binney92}, 
\citet{ostriker and binney89}
studied how a disk of massive rings reacts to a slewing disk potential.
They found that the reaction of the disk depends largely on its surface
density, with warps appearing in regions of low surface density.
Also motivated by the cosmic infall of material with differing
angular momentum, \citet{debattista and sellwood99}
found that when the angular momentum of a halo and disk are misaligned,
dynamical friction transfers angular momentum between them in such
a way as to produce a long-lived warp.

More recently, \citet{lopez-corredoira et al02a} examined the torques
produced by the transfer of angular momentum from gas falling onto a
galactic disk, and the warping of the disk in response to
these torques as well as the internal torques due to the interaction
of different rings within the disk.
They find a disk response which mirrors the
Milky Way warp, though the effects of embedding the disk in a dark
halo are as yet insufficiently modelled.

In a cosmological setting, a galactic disk experiences
gravitational torques
from three different sources: external galaxies, non-spherically-symmetric
substructure in the halo, and misalignment between the disk and halo;
if the dark matter halo is not spherical, as suggested both by simulations
\citep[e.g.][]{fwed85,katz91,dubinski and carlberg91,cole and lacey96} and
observations \citep{sackett99}, then it will exert a gravitational torque
on a disk not in its symmetry plane.

In this paper, we evaluate the effect of the gravitational tidal torques
a typical galactic disk experiences from its misalignment with
the halo, and study whether these torques provide a possible origin
for warped disks. We first develop a framework for how orbits in a
disk react to torques. We then investigate the form and magnitude
of torques due to flattened misaligned halos. The effect of these
torques on massless disks is examined by analytic and numerical techniques.
Finally, we perform N-body simulations of massive disks resembling
the Milky Way and investigate how the self-gravity of the disk may
affect its reaction to cosmological torques.

\section{Origin of Torques on Galactic Disks}

\subsection{Framework for Torqued Orbits\label{const radius orbit}}

\subsubsection{Introduction to torque}
\begin{figure}
\input{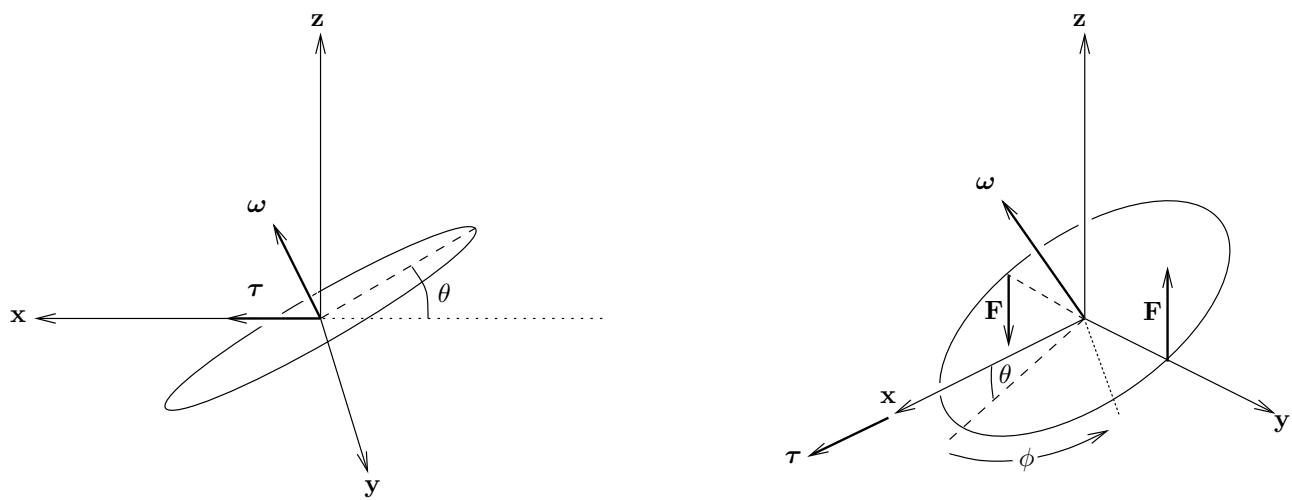}

\caption{
  Two views of a circular orbit with angular velocity vector
    ${\bolder{\omega}}$ tilted from the $xy$-plane toward the $x$-axis
    by an angle $\theta$. $\phi$ is the azimuthal angle around
    the orbit. A torque ${\bolder{\tau}}$ directed along the $x$-axis
    produces forces ${\bolder{F}}$, as in equation~(\ref{ai is Tij rj}). 
    \label{angle diagram}
}
\end{figure}

The detailed reaction of a stellar disk to a torque is derived in
section~\ref{orbit section}. Here we introduce a simplified model to establish
a framework in which the reaction of the disks to the torques can
be studied, and define the terms and symbols we use.
Consider a star in the disk on a circular orbit in the
\( xy \)-plane around the origin with angular velocity \( \omega (r)=v_{c}(r)/r \)
(see Figure~\ref{angle diagram}). We apply a torque around the \( x \)-axis
by accelerating the star in the \( y \) and \( z \) directions around
the \( x \)-axis with an antisymmetric pseudo-tensor \( T_{ij} \).
If the acceleration in the \( i \) direction is \( a_{i} \) and
the \( j \)-position of the star is \( r_{j} \) then the applied
acceleration is\begin{equation}
\label{ai is Tij rj}
a_{i}=\sum _{j}T_{ij}r_{j}
\end{equation}
where\[
{\bolder T}=\left( \begin{array}{ccc}
0 & 0 & 0\\
0 & 0 & -\tau (r)\\
0 & \tau (r) & 0
\end{array}\right) .\]
This is an angular acceleration (or equivalently a specific torque)
around the \( x \)-axis of magnitude \( a/r =\tau (r) \).
Since the angular acceleration is the time derivative of the angular
velocity, the \( x \)-component of the angular velocity increases
in response to the torque,
and the angular momentum of the star will slew toward the \( x \)
axis, tilting the orbit toward the $x$-axis.

\subsubsection{Direction of tilt}
\label{direction of tilt}
\begin{figure}
\plotone{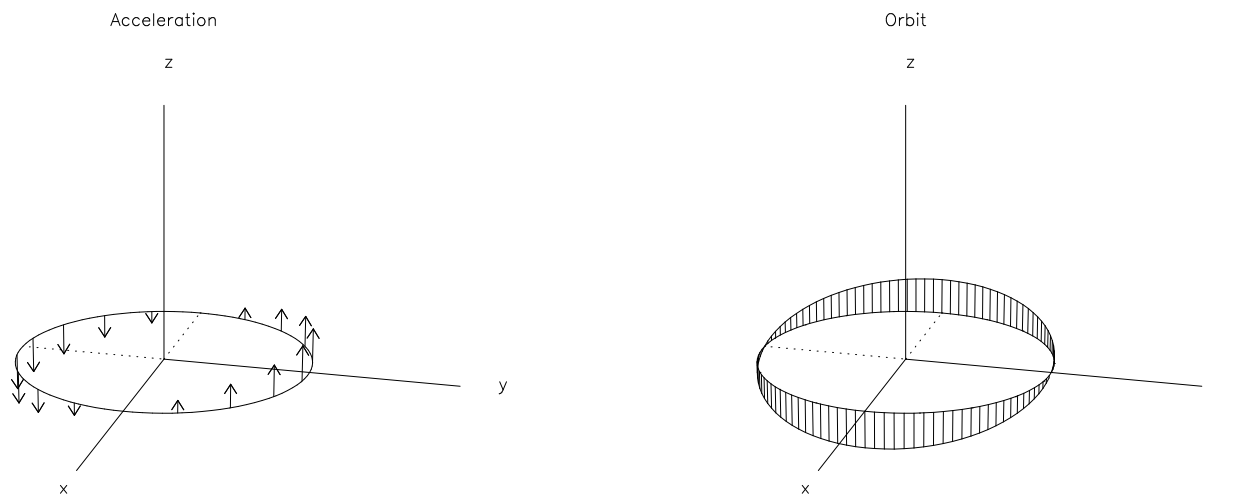}

\caption{
  {\em Left:} Acceleration vectors due to the
   torque for an orbit in the $xy$-plane.
  {\em Right:} The $z$ response lags $y$ by one quarter
   of an orbit, causing the orbit to tilt around the $y$
     axis toward the $x$ axis.
  \label{risen orbits}
}
\end{figure}

It is worth taking a minute here to elaborate on the point that the
orbit tilts toward the $x$ axis. One might think that a torque about
the $x$ axis would tilt the plane of the orbit \textit{around} the
$x$ axis, rather than toward it.
Indeed, this is what happens initially,
and seems to be the thinking behind
attempts to explain the Milky Way warp using the Magellanic
Clouds \citep[e.g.][]{weinberg98,tsuchiya02}.
However, \citet{garcia-ruiz et al00} examined the Lagrangian of
the system and conducted N-body experiments to demonstrate that 
a warp caused by the Magellanic Clouds would be perpendicular to
the observed Milky Way warp.

This result can be understood by examining the equations of motion
of the system for a particle on a circular orbit in the $xy$ plane
about the $z$-axis:
\begin{equation}
\frac{d^2 \bolder{r}}{dt^2} = \frac{1}{r} \frac{\partial\phi}{\partial r}
	\bolder{r} + \bolder{\tau} \times \bolder{r}.
\end{equation}
If the untorqued orbit has an angular frequency~$\omega$, and the
torque is directed around the $x$-axis, then the equations of motion
for the three Cartesian coordinates are
\begin{eqnarray}
 \frac{d^2 x}{dt^2} &=& -\omega^2 x \\
 \frac{d^2 y}{dt^2} &=& -\omega^2 y - \tau z \label{ydd eq} \\
 \frac{d^2 z}{dt^2} &=& -\omega^2 z + \tau y. \label{zdd = -w2 z + t y}
\end{eqnarray}
The $x$ motion is decoupled from the other coordinates, and therefore is
unaffected by the torque.
By substituting equation~(\ref{ydd eq}) into equation~(\ref{zdd = -w2 z + t y}),
the system of equations can be converted
into a single fourth-order ordinary differential equation for $z$ of the form
\begin{equation}
\frac{d^4 z}{dt^4} + 2\omega^2 \frac{d^2 z}{dt^2} + (\omega^4+\tau^2)z = 0.
\end{equation}
Using the ansatz $z=A e^{i\Omega t}$,
  the roots of the characteristic equation are
\begin{equation}
\Omega^2 = \omega^2 \pm i\tau.
\label{freq Omega}
\end{equation}
The two solutions correspond to growing ($-$) and decaying ($+$) modes.
The growing mode will dominate over time, so we neglect the decaying mode.
As $d^2 z/dt^2 = - \Omega^2 z$, equations~(\ref{zdd = -w2 z + t y})
 and~(\ref{freq Omega}) yield
\begin{equation}
y = i z.
\end{equation}
For an oscillating $y$, this means that the phase of $y$ is a quarter
of an orbit ahead of $z$, or that the phase of $z$ is a quarter of an
orbit behind $y$.
As demonstrated in Figure~\ref{risen orbits}, $z$ reaches
its maximum a quarter of an orbit after $y$, causing the orbit to
tilt around the $y$-axis toward the $x$-axis.

  Therefore, we expect that any tilting (and
    therefore warping, which is a tilt that varies with radius)
    will occur toward the plane perpendicular to the torque.
Since a torque is a transfer of angular momentum, a simple way to understand
the result of \citet{garcia-ruiz et al00} is that the galactic disk
acquires angular momentum with the same direction as the orbital angular
momentum of the satellite, and therefore it will warp toward the plane
of the satellite's orbit, not toward the satellite itself.

\subsubsection{Tilting timescale}
For an orbit in the \( xy \)-plane, the angular momentum is initially
aligned with the \( z \) axis (\( {\bolder \omega }=\omega _{0}(r)\,  \)\uvec{z}).
The torque adds angular momentum in the \( x \) direction at a constant
rate:\[
\frac{d\bolder {\omega }}{dt}=\tau (r)\, \uvec {x}.\]
If the radius of the orbit stays constant, then the angular velocity
grows as\[
\omega (t)=t\tau (r)\, \uvec {x}+\omega _{0}(r)\, \uvec {z}.\]

As the torque adds angular momentum along the \( x \) axis, the the
orbit tilts toward the \( x \) axis by an angle \( \theta  \), where
\begin{equation}
\label{simple theta(t)}
\tan \theta =\frac{\omega _{x}}{\omega _{z}}=\frac{t\tau (r)}{\omega _{0}(r)}.
\end{equation}
 The rate of tilting at small angles is\[
\frac{d\theta }{dt}=\frac{\tau (r)}{\omega _{0}(r)}\]
with a characteristic timescale to tilt one~radian of \begin{equation}
\label{ttilt is omega over tau}
t_{\mathrm{tilt}}=\frac{\omega _{0}(r)}{\tau (r)}.
\end{equation}

Another motivation for this form is that \( \omega _{0} \) is the
angular momentum per unit mass and \( \tau  \) is the specific torque,
which is the rate of change of
 the specific angular momentum. Note that even
for a flat rotation curve where \( \omega _{0}=v_{c}/r \), \( t_{\mathrm{tilt}} \)
is a monotonically increasing function of \( r \) if \( \tau (r) \)
falls off faster than \( r^{-1} \); i.e.~for a torque
that decreases with radius, the inner regions of the
disk should always tilt faster than the outer regions, resulting in
a trailing warp.

\subsection{Expected torques from misaligned flattened halos\label{idealized torques}}

There is observational \citep{sackett99} and theoretical \citep{katz91,cole and lacey96}
evidence that dark matter halos are flattened with \( c/a \) axis
ratios ranging as low as 0.5. Cosmological simulations 
\citep{fwed85,katz91,warren et al92,porciani et al02}
show that while the angular momentum and minor axis of a flattened
halo will be correlated, there will be significant misalignments between
them in a large number of halos.
Recent high-resolution simulations suggest that the angular momentum
of the baryons may not even be aligned with that of the dark matter
\citep{vdb et al02}, and that therefore the disks that form from those
baryons are usually misaligned
with the minor axis of the halo matter distribution.

We calculate the magnitude of the torque for a disk misaligned in
a flattened halo with a radial dependence of an NFW profile \citep{NFW96}
but flattened along the \( z \) axis:
\begin{equation}
\label{flattened rho NFW}
\rho (x,y,z)=\frac{\rho _{0}}{\left( m/r_{s}\right) \left( 1+m/r_{s}\right) ^{2}}
\end{equation}
for a modified radius \( m^{2}=x^{2}+y^{2}+z^{2}/q^{2} \) where \( q \)
is the \( c/a \) axis ratio. The force from this distribution on
a given point is calculated using equation (2-88) of \citet{BT}.
For reference, we also calculate the torques inside flattened isothermal
profiles, as in equation (2-54a) of \citet{BT}. The torques are calculated
by evaluating the forces at opposite points along a fictitious disk
centered at the origin and placed at an angle \( \theta  \) to the
\( xy \) plane (see Figure~\ref{angle diagram}).
The acceleration at radius \( r \) due to a torque
of magnitude \( \tau  \) is out of the plane, is of opposite sign
on opposite sides of the disk, and is of magnitude \( F=\tau r \).
Given the forces on two test points \( F_{1} \) and \( F_{2} \)
at opposite sides of the disk, we take the component of the force
out of the plane of the disk \( F_{\bot 1} \) and \( F_{\bot 2} \).
The torque is \[
\tau =\frac{F_{\bot 1}-F_{\bot 2}}{2r}.\]
 The angular velocity \( \omega  \) is determined at each point from
the radial force: \( \omega =\sqrt{F_{r}/r} \) where the radial force
\( F_{r} \) is defined to be positive when it is directed
toward the center of the halo. The average value of \( \omega  \)
for the two points across the disk is used. The tilting timescale
is given by equation~(\ref{ttilt is omega over tau}).

\begin{figure}
\plotone{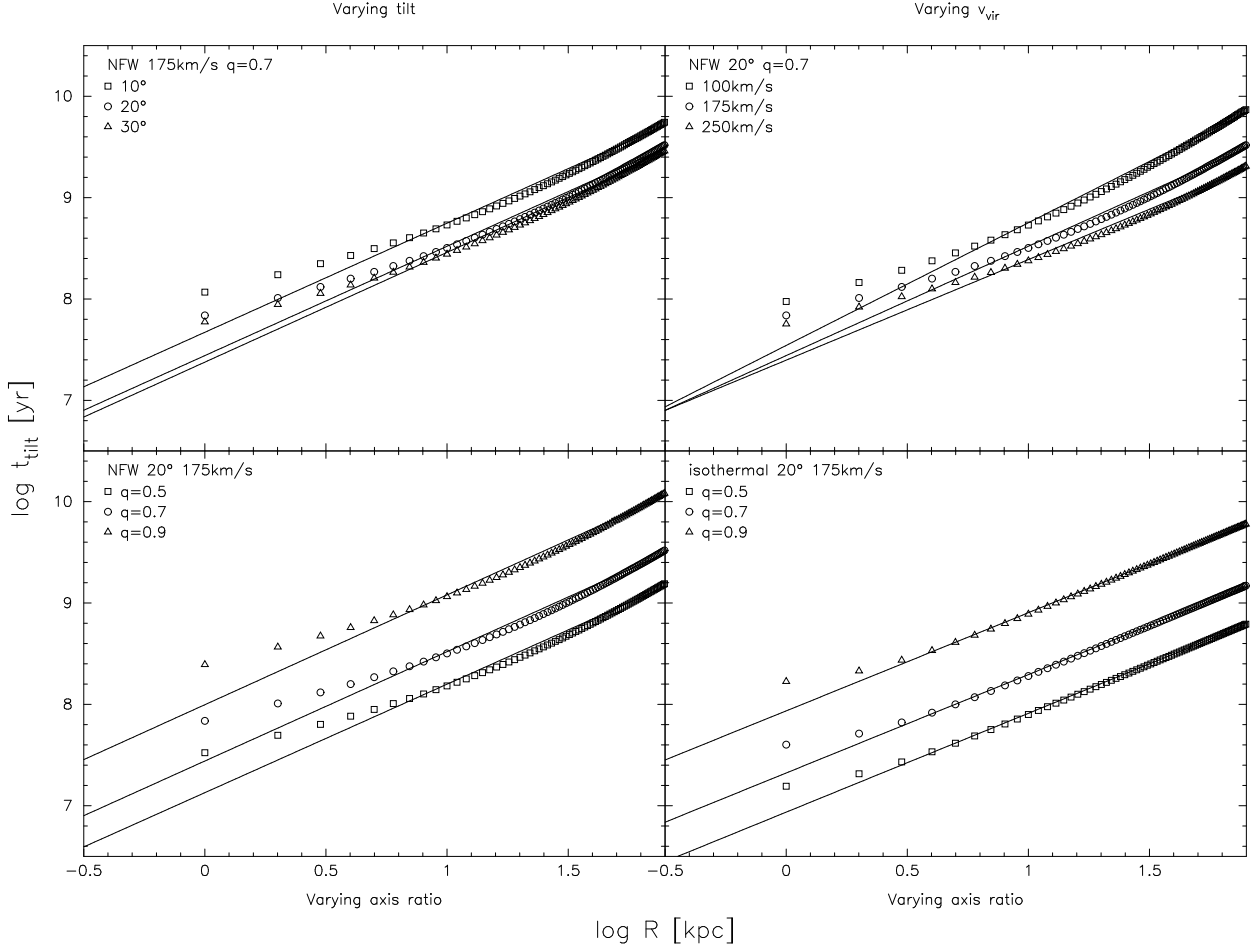}

\caption{
The symbols indicate the torques experienced by disks misaligned
inside flattened halo profiles, expressed in terms of the tilting
timescale $t_{\mathrm{tilt}}$.
The lines are power law fits for \protect\( \log R>0.1\protect \).
Note that stronger torques have \emph{shorter} timescales and appear
lower in these graphs.
\textit{Top-left:} Torques inside NFW halos with virial velocities
$v_{200}=175\mathrm{\, km\, s^{-1}}$, flattenings $q=0.7$,
and concentration parameters $c_{200}=15$,
for disks misaligned by 10\degr,~20\degr, and 30\degr\ from the symmetry plane.
\textit{Top-right:} Torques inside NFW halos with virial velocities
$v_{200}=100,~175,$ and~$200\mathrm{\, km\, s^{-1}}$,
flattenings $q=0.7$,
and concentration parameters $c_{200}=15$,
 for disks misaligned by 20\degr\ from the symmetry plane.
\textit{Bottom-left:} Torques inside NFW halos with virial velocities
$v_{200}=175\mathrm{\, km\, s^{-1}}$,
flattenings $q=0.5,~0.7,$ and~$0.9$,
and concentration parameters $c_{200}=15$,
for disks misaligned by 20\degr\ from the symmetry plane.
\textit{Bottom-right:} Torques inside isothermal halos with virial velocities
$v_{200}=175\mathrm{\, km\, s^{-1}}$,
flattenings $q=0.5,~0.7,$ and~$0.9$,
and core radii \protect\( R_{c}=1\protect \)~kpc,
for disks misaligned by 20\degr\ from the symmetry plane.\label{vary figure}}
\end{figure}

Figure~\ref{vary figure} illustrates the torques derived using this method
for halos with a cross-section of properties expected for galactic halos.
The fiducial NFW model has flattening
\( q=0.7 \), virial velocity \( v_{c}=175\mathrm {\, km\, s^{-1}} \)
  (corresponding to $v_{\mathrm{rot}}=205\mathrm{\, km\, s^{-1}}$
   at a galactocentric radius of 10~kpc),
and a disk at an angle 20\degr\ to the symmetry plane of the halo. All NFW
halos have concentration parameters \( c_{200}=15 \) \citep{NFW97}.
The different panels
show the effect of varying the disk angle by 10\degr, the axis ratio
from 0.5~ to~0.9, and the virial velocity by 75~km~s\( ^{-1} \)
compared to the fiducial model.
The bottom-right panel of Figure~\ref{vary figure} shows the torques
inside an isothermal profile with a range of axial ratios. The magnitude
of the torque in the isothermal case is similar to that in the NFW
case, but falls off less rapidly with radius.

Figure~\ref{vary figure} demonstrates that the torque from a misaligned
flattened halo is significant to its evolution on a cosmological timescale,
as the tilting timescale \( t_{\mathrm{tilt}} \) is less then the
Hubble time over the entire length of the disk for all halos modelled.
The torques scale approximately as the density, and therefore
  fall off with radius as \( \tau (r)\propto r^{\beta } \)
with \( \beta  \) ranging from \( -1 \) at small radii to \( -3 \)
at large radii for NFW halos and equal to \( -2 \) for isothermal
halos. The typical torque follows the relation\begin{equation}
\label{tau = 1e-30 r-2.5}
\tau (r)=10^{-30}\left( \frac{r}{1\, \mathrm{kpc}}\right) ^{-2.5}\, \mathrm{s}^{-2}.
\end{equation}
 The angular velocity \( \omega (r)\propto r^{\alpha } \) where \( \alpha  \)
ranges from \( -0.5 \) to \( -2 \) in NFW halos, and is \( -1 \)
in isothermal halos.
The tilting timescale \( t_{\mathrm{tilt}} \), as calculated
using equation~(\ref{ttilt is omega over tau}),
rises almost linearly with radius; the timescale is shorter in
the inner regions of the disk, and therefore the inner disk tilts
faster than the outer disk in response to torques from misaligned
halos.
For example,
in an isothermal halo $\tau(r) \propto r^{-2}$ while 
$\omega(r) \propto r^{-1}$ so
$t_{\mathrm{tilt}} = \omega(r)/\tau(r) \propto r$.
The torque increases more rapidly toward the center than does the
disk's ability to resist the torque due to its angular momentum.

The further the halos and disks are misaligned, the stronger the torque
is; however, the torque profile is mostly unchanged for angles beyond
20\degr. Increasing the virial velocity, and therefore the mass of
the halo, increases the magnitude of the torque. Because the concentration
\( c_{200} \) is held constant amongst these models, changing the
virial velocity also changes the scale radius \( r_{s} \), which
can be seen in the shifting of the radius of the knee in the profiles
of the top-right panel of Figure~\ref{vary figure}. The flattening
of the halo has a large effect on the magnitude of the torque, with
torques strengthening as the halo departs further from spherical symmetry.
The torques fall off more slowly in the isothermal halos than in the
NFW halos, but are of similar magnitude over most of the disk radius.

\section{Orbits in Torqued Disks\label{orbit section}}

\subsection{Reaction of a massless disk\label{massless disk full derivation}}

The results of section~\ref{const radius orbit} assume that the
torque does not change the radius of the orbit. This is a poor approximation,
since the torque does work on the star whenever the torque is not
orthogonal to the orbital angular momentum. In this section we do
a more thorough analysis, taking into account the changing radius
of the orbit.

We define \( \theta  \) as in section~\ref{const radius orbit},
and \( \phi  \) as the azimuthal angle around the orbit (see Figure~\ref{angle diagram}).
Based on the results of section~\ref{idealized torques}, the torque
can be expressed as a power law:\begin{equation}
\label{tau definition}
{\bolder \tau }=\tau _{0}\left( \frac{r}{r_{0}}\right) ^{\beta }\, \uvec {x}
\end{equation}
which exerts a force\begin{equation}
\label{F definition}
\begin{array}{rcl}
{\bolder F} & = & m{\bolder \tau }\times {\bolder r}\\
 & = & m\tau _{0}r_{0}\left( \frac{r}{r_{0}}\right) ^{\beta +1}(\cos \phi \, \sin \theta \, \uvec {y}+\sin \phi \, \uvec {z}).
\end{array}
\end{equation}

This force torques the orbit and alters the angular momentum \( \bolder {L} \).
If the orbit can be considered to be a circular orbit with circular
velocity \( v_{c}(r) \) (where the radial dependence is a function
of the underlying potential), then

\begin{equation}
\label{L definition}
{\bolder L}={\bolder r}\times {\bolder p}=mrv_{c}(r)\, \uvec {\omega }
\end{equation}
where \( \uvec {\omega } \) is the unit vector in the direction perpendicular
to the orbit. \( \bolder {L} \) has an additional component of order
\( mr^{2}(d\theta /dt)\uvec {y} \) because the orbit is tilting,
but it does not provide an independent constraint,
so we neglect it.
The time derivative of \( \bolder {L} \)
is\begin{equation}
\label{dL/dt definition}
\begin{array}{c}
\frac{d{\mathbf{L}}}{dt}=\left[ mv_{c}(r)\sin \theta \frac{dr}{dt}+mr\sin \theta \frac{dv_{c}}{dr}\frac{dr}{dt}+mrv_{c}(r)\frac{d\theta}{dt}\, \cos \theta \right] \: \uvec {x}\\
+\left[ mv_{c}(r)\cos \theta \frac{dr}{dt}+mr\cos \theta \frac{dv_{c}}{dr}\frac{dr}{dt}-mrv_{c}(r)\frac{d\theta}{dt}\, \sin \theta \right] \: \uvec {z}.
\end{array}
\end{equation}

The change in angular momentum comes from the torque
${\bolder{N}}$ imparted to the
system: \( \bolder {N=r\times F} \). Taking \( \bolder {F} \) from
equation~(\ref{F definition}) and averaging over one orbit to remove the $\phi$
dependence (assuming that \( dt\propto d\phi  \) over one orbit, i.e.
that \( \omega (r) \) is slowly varying so the \( \phi  \)-averaged
torque is equal to the time-averaged torque) gives\begin{equation}
\label{<N>}
\begin{array}{c}
\left\langle \bolder {N}\right\rangle =\frac{1}{2}m\tau _{0}r_{0}^{2}\left( \frac{r}{r_{0}}\right) ^{\beta +2}(1+\sin ^{2}\theta )\, \uvec {x}\\
+\frac{1}{2}m\tau _{0}r_{0}^{2}\sin \theta \, \cos \theta \left( \frac{r}{r_{0}}\right) ^{\beta +2}\, \uvec {z}
\end{array}
\end{equation}
Setting \( d\bolder {L}/dt=\left\langle \bolder {N}\right\rangle  \)
and solving for \( dr/dt \) and \( d\theta /dt \),\begin{eqnarray}
\frac{dr}{dt} & = & \tau _{0}r_{0}^{2}\left( \frac{r}{r_{0}}\right) ^{\beta +2}\left[ v_{c}(r)+r\frac{dv_{c}}{dr}\right] ^{-1}\sin \theta \label{dr/dt coupled} \\
\frac{d\theta }{dt} & = & \frac{1}{2}\frac{\tau _{0}r_{0}}{v_{c}(r)}\left( \frac{r}{r_{0}}\right) ^{\beta +1}\cos \theta .\label{dth/dt coupled} 
\end{eqnarray}

We could not find an analytic solution to this coupled pair of differential
equations, so they need to be solved numerically. However, if \( dr/dt \)
is small then equation~(\ref{dth/dt coupled}) can be solved at a fixed radius.
The solution is\begin{equation}
\label{analytic better theta(r,t)}
\theta (r,t)=2\cos ^{-1}\left[ \frac{1}{\sqrt{2}}\frac{1+e^{t/t_{0}(r)}}{\sqrt{1+e^{2t/t_{0}(r)}}}\right] 
\end{equation}
where the timescale at a given radius is\begin{equation}
\label{timescale for theta(r,t)}
t_{0}(r)=\frac{2v_{c}(r)}{\tau _{0}r_{0}}\left( \frac{r}{r_{0}}\right) ^{-\beta -1}.
\end{equation}
\begin{equation}
= 2 \frac{\omega_0(r)}{\tau(r)}
\label{timescale is half omega over tau}
\end{equation}
and the circular velocity \( v_{c}(r) \) depends on the halo potential.

Comparison of equations~(\ref{timescale is half omega over tau}) 
and~(\ref{ttilt is omega over tau}) reveals that the main effect of allowing
a second degree of freedom, by permitting $r$ to vary, is to increase the
tilting timescale by a factor of two. 
This factor is also evident from a comparison
of the dashed and the solid line in Figure~4.
Theoretical studies of warps which
artificially restrict this degree of freedom, for example by
 approximating the disk
of stars by solid rings of constant radius,
systematically overestimate the effect of the torques on the disk by
a factor of two.
\label{factor of 2 derivation}

\subsection{Comparison with simulations\label{analytic vs simulations}}

\begin{figure}
\plotone{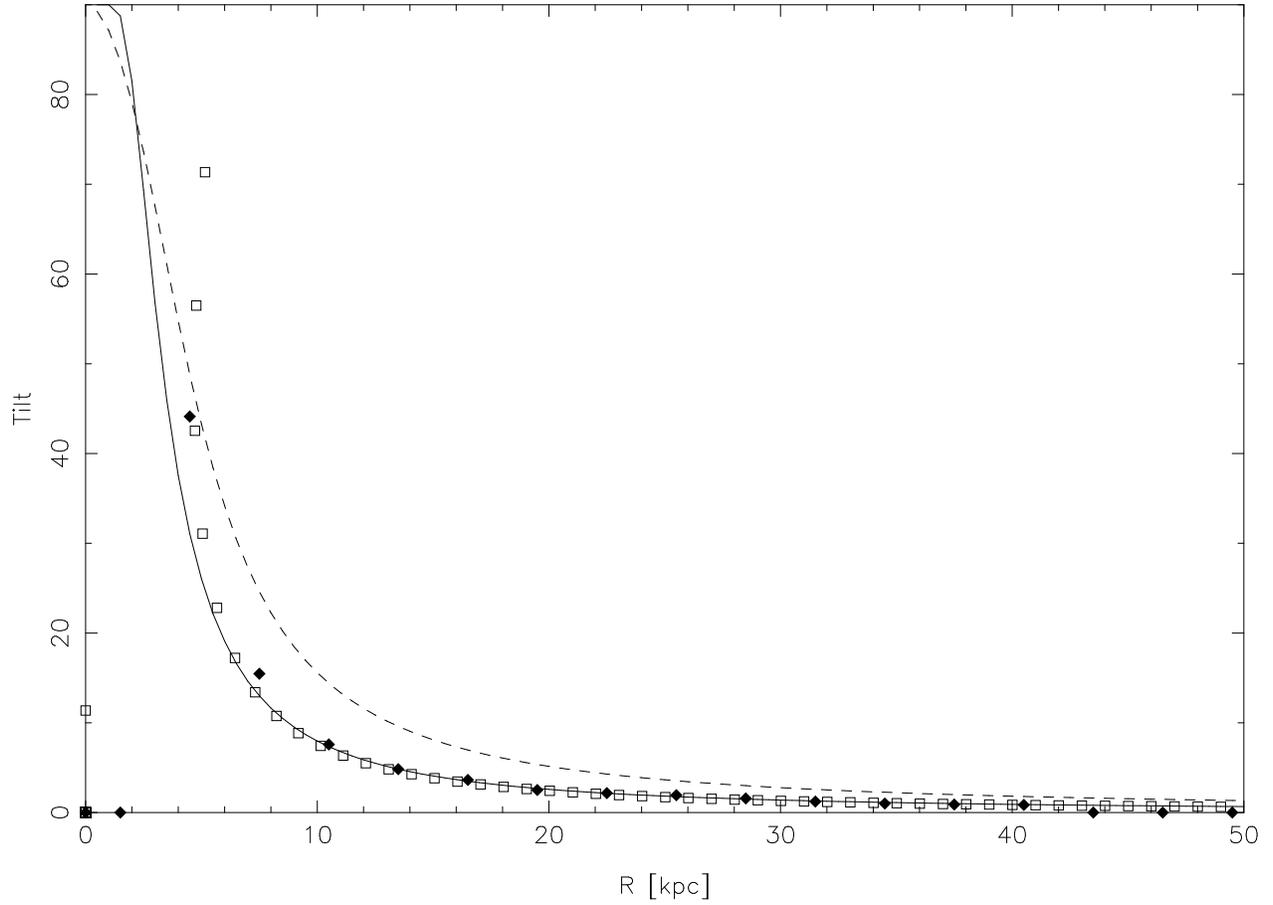}

\caption{Tilt versus radius for rings in a disk subject to a torque of the
form (\ref{tau definition}) with \protect\( \beta =-2.5\protect \)
and \protect\( \tau _{0}=10^{-30}\; \mathrm{s}^{-2}\protect \) at
\protect\( r_{0}=1\; \mathrm{kpc}\protect \) for 1~Gyr. Filled diamonds
correspond to a N-body simulation (binned in spherical shells 3~kpc wide),
open squares to the numerical solution of equations~(\ref{dr/dt coupled})
and~(\ref{dth/dt coupled}), the dashed line to the analytic
prediction from the simplified
model of equation~(\ref{simple theta(t)}), and the solid
line to the analytic prediction of equation~(\ref{analytic better theta(r,t)}).\label{plot massless theta(r,t)}}
\end{figure}
The simplified model of equation~(\ref{simple theta(t)}),
the detailed derivation
(\ref{analytic better theta(r,t)}), and the numerical solution to
equations~(\ref{dr/dt coupled}) and~(\ref{dth/dt coupled}), provide
three predictions of how a torqued orbit should evolve.
To see how well these predictions work, we compare them to a simulated
disk of massless particles subject to a torque that takes all the forces
explicitly into account.

We perform a simulation of a disk of 16000~massless tracer particles placed
in 800~concentric rings at radii spaced 0.05~kpc apart, each containing
20~particles. They are evolved for 1~Gyr using a predictor-corrector
code under the influence of an NFW potential (\( c_{200}=15 \), \( r_{200}=200\; \mathrm{kpc} \))
and a torquing force of the form~(\ref{tau = 1e-30 r-2.5}).
We cap the torque at $\tau = \tau_0$ inside $r_0$ in order to prevent
extremely small timesteps for particles near the center where
the torquing force diverges.
The particles were then binned into spherical shells 3~kpc wide.
We calculated the moment of inertia tensor \( I_{ij}=\sum _{k}m_{k}x_{i,k}x_{j,k} \)
for the particles within each shell and diagonalized it to find the
direction of the minor axis (the axis with the smallest moment of
inertia). The minor axis was found to be tilted toward the \( x \)-axis
by an angle \( \theta  \) that varied from bin to bin.

We also evolved the coupled set of differential equations~(\ref{dr/dt coupled})
and~(\ref{dth/dt coupled}) forward 1~Gyr with the same values of
\( \beta  \) and \( \tau _{0}(r_{0}) \) as above using a fourth-order
Runge-Kutta integrator for a set of initial conditions \( \theta =0 \),
\( r=0,0.5,1,1.5\ldots 100\; \mathrm{kpc} \) representing an initially
flat disk.

Figure~\ref{plot massless theta(r,t)} compares the simulation (filled
diamonds) with the numerical solution to equations~(\ref{dr/dt coupled})
and~(\ref{dth/dt coupled}) (open squares), the simplified
model~(\ref{simple theta(t)}) (dashed line), and the detailed analytic 
prediction~(\ref{analytic better theta(r,t)}) (solid line).
It is apparent that at most radii, (\ref{analytic better theta(r,t)})
is a good approximation to the simulation --- much better 
than~(\ref{simple theta(t)}), demonstrating the importance of including
the radial drift of the orbits.
At small tilt angles, the difference between
the solid line~(\ref{analytic better theta(r,t)}) and
the dashed line~(\ref{simple theta(t)})
 is exactly the factor of~two due to the extra degree of freedom
derived in Section~\ref{factor of 2 derivation}.
At small radii (\( r\la 5\: \textrm{kpc} \)),
the solutions diverge: the
numerical solution contains rings of the same radius but different
tilts while the simulation bins the particles by radius and therefore
averages particles in the same shell together; more importantly, the
assumption that \( dr/dt \) is small is violated. However, in the
innermost region of a halo the symmetric bulge will dominate the potential,
and therefore the torque power law is likely invalid in any case.
For these reasons, we ignore the innermost 2~kpc in our analysis
and consider 
equation~(\ref{analytic better theta(r,t)}) to be a good description
of the disk everywhere else.

\section{Self-gravitating Disks and Warps}

Figure~\ref{plot massless theta(r,t)} suggests that disks should
have continuously curved warps in their inner regions, eventually
becoming flat in the outer regions where \( t_{\mathrm{tilt}} \)
is small. Real warped galaxies appear quite the opposite; they are
solid out to approximately the Holmberg radius (half of the diameter
of the \( I=26.5\, \mu _{\mathrm{pg}} \) isophote) and warped beyond
that \citep{briggs90}. The most important difference between real
galactic disks and those used in section~\ref{orbit section} is
that the self-gravity of real galactic disks is important; galaxies
have mass. 

Although the dark matter halo must dominate the potentials of disk
galaxies at large radii, there is considerable evidence that the self-gravity
of the disk is important to its dynamics over much of its area. \citet{broeils and courteau97}
show that {}``maximal disk'' models, where the mass-to-light ratios
of the disks \( (M/L) \) are as large as is consistent with the rotation
curves, provide good fits to most rotation curves.
\citet{sackett97} demonstrates that the Milky Way is also
consistent with having a maximal disk. Although
\citet{courteau and rix99} have used the residuals
of the Tully-Fisher relation \citep{tully fisher} to
estimate that disks provide only 40\% of the dynamical mass at 2.2
exponential scale lengths,
and  \citet{de blok and mcgaugh97} have demonstrated that
the rotation curves of low surface brightness galaxies imply that
they are substantially sub-maximal,
the prevalent phenomena of spiral arms and bars, which are disk instabilities,
require that in many disks, the self-gravity
of the disk must be important to its dynamics \citep{athanassoula et al 87}.

The self-gravity of the disk acts to keep the disk flat. \citet{ostriker and binney89}
examined the effect of a slewing disk potential on a set of self-gravitating
rings and found that regions of high surface density react like a
solid body. Since the surface density of the disk is highest in the
central regions, the central parts of the disk will be kept locally
flat, resulting in disks that closer resemble observed warped galaxies.
Using numerical N-body simulations, we investigate this effect.

\citet{lopez-corredoira et al02a} take this into account by developing a
form for the internal torque $\tau_{\mathrm{int}}$. Here we do not use
their $\tau_{\mathrm{int}}$, but rather integrate the 
orbits numerically. While this does not provide us with an explicit
set of differential equations for $\theta(r,t)$, it frees us from such
assumptions as the constancy of $\omega(r)$ (which we saw was
important in Section~\ref{massless disk full derivation}), the
existence of an equilibrium configuration, and the lack of a dark
matter halo.

\subsection{The simulations}

To evaluate how the mass of the disk affects the results of section~\ref{orbit section},
we performed numerical N-body simulations of massive Milky Way type
galactic disks. Equilibrium disk models of disk mass \( 1\times 10^{10}\, \mathrm{M}_{\sun } \),
\( 3\times 10^{10}\, \mathrm{M}_{\sun } \), and \( 5.6\times 10^{10}\, \mathrm{M}_{\sun } \),
scale length \( r_{d}=3.5\, \mathrm{kpc} \) and vertical scale height
\( h_{z}=325\, \mathrm{pc} \) in a static spherically-symmetric NFW
halo potential \citep{NFW97} with concentration parameter \( c_{200}=15 \)
and virial velocity \( v_{200}=175\, \mathrm{km}\, \mathrm{s}^{-1} \)
were constructed using the method of \citet{hernquist93} and then
allowed to relax under the force of gravity until they appeared to
be in equilibrium. The halo masses were varied such that the mass
of the disk plus halo was \( 4.1\times 10^{11}\: M_{\sun } \) in
each case.

The models were evolved using the GRAPESPH code \citep{steinmetz96},
where interparticle gravitational forces are computed using direct
summation with GRAPE-3 hardware \citep{grape3}. A plummer softening
of 0.3~kpc has been used. The models were evolved for 2~Gyr, which
took 5000-7000 timesteps depending on the model. The torque force
was applied as an external force
as described in section~\ref{analytic vs simulations}
with \( \tau _{0}=10^{-30}\: \mathrm{s}^{-2} \) at \( r_{0}=1\: \mathrm{kpc} \).
Simulations were performed with torque slopes of \( \beta =-2.0, \)
\( -2.5, \) and \( -3.0 \).

The disks contained 16384 particles. The \( 3\times 10^{10}\: M_{\sun } \)
simulations were also performed with 32767 particles to see if the
resolution was sufficient. The results for the high resolution simulations
were identical to those for the lower resolution simulations to well
within the angular errors computed in section~\ref{warp evolution section}
except for in the spherical shells which contained very few particles
and had been neglected from the analysis on those grounds.

\subsection{Warped disks}

\begin{figure}
\plotone{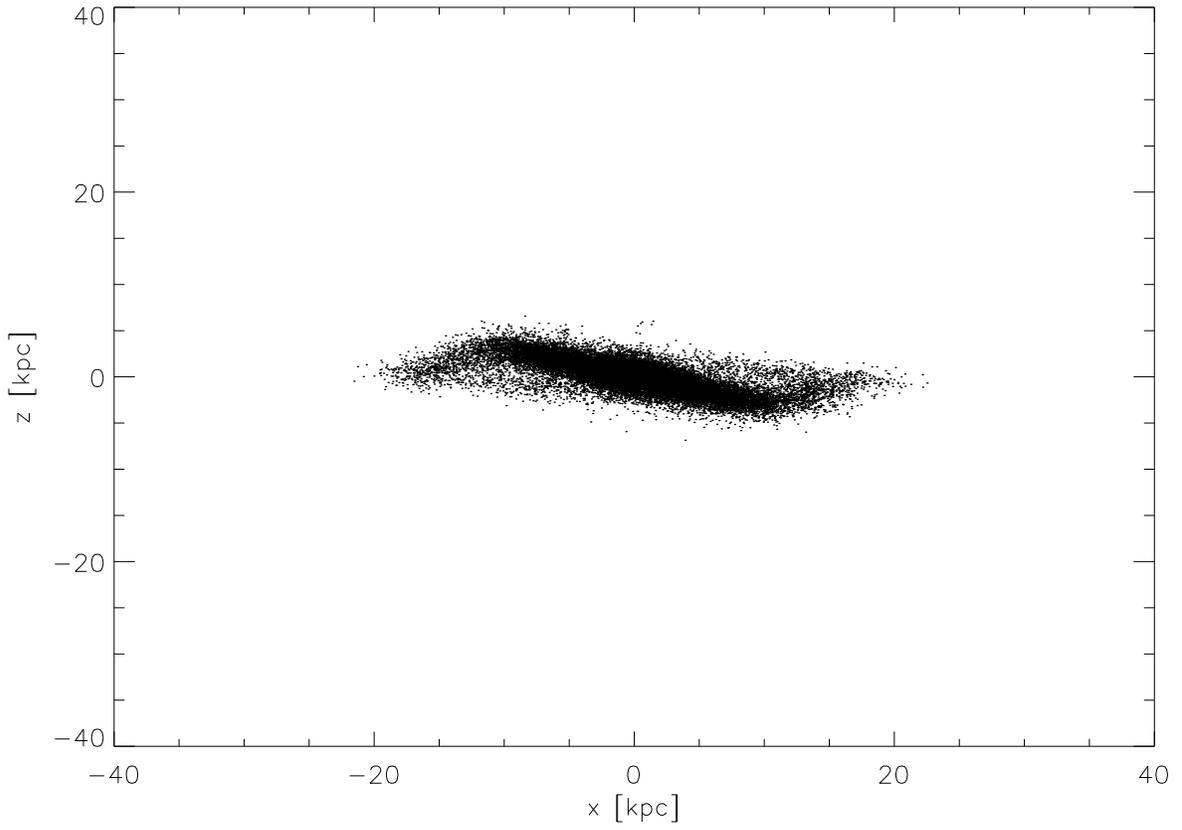}

\caption{$x$-$z$ projection of a simulated disk galaxy
  with mass \protect\( 3\times 10^{10}\: M_{\sun }\protect \)
after 1~Gyr in a torque of \protect\( \tau _{0}=10^{-30}\: \mathrm{s}^{-2}\protect \)
at \protect\( r_{0}=1\: \mathrm{kpc}\protect \) and \protect\( \beta =-2.5\protect \).\label{simulation disk}}
\end{figure}
Figure~\ref{simulation disk} shows the simulation of a \( 3\times 10^{10}\: M_{\sun } \)
disk with 32767 particles subject to a \( \beta =-2.5 \) torque for
1~Gyr. The main plane of the disk is flat and clearly tilted toward
the positive \( x \)-axis,
as expected from Section~\ref{direction of tilt}.
  Beyond 10~kpc, the disk no longer remains
flat but warps back toward the original plane. The particles which
appear to be filling in the area between the main disk and the warp
are projections and are actually in front of or behind the galaxy
at large radii.

\begin{figure}
\plotone{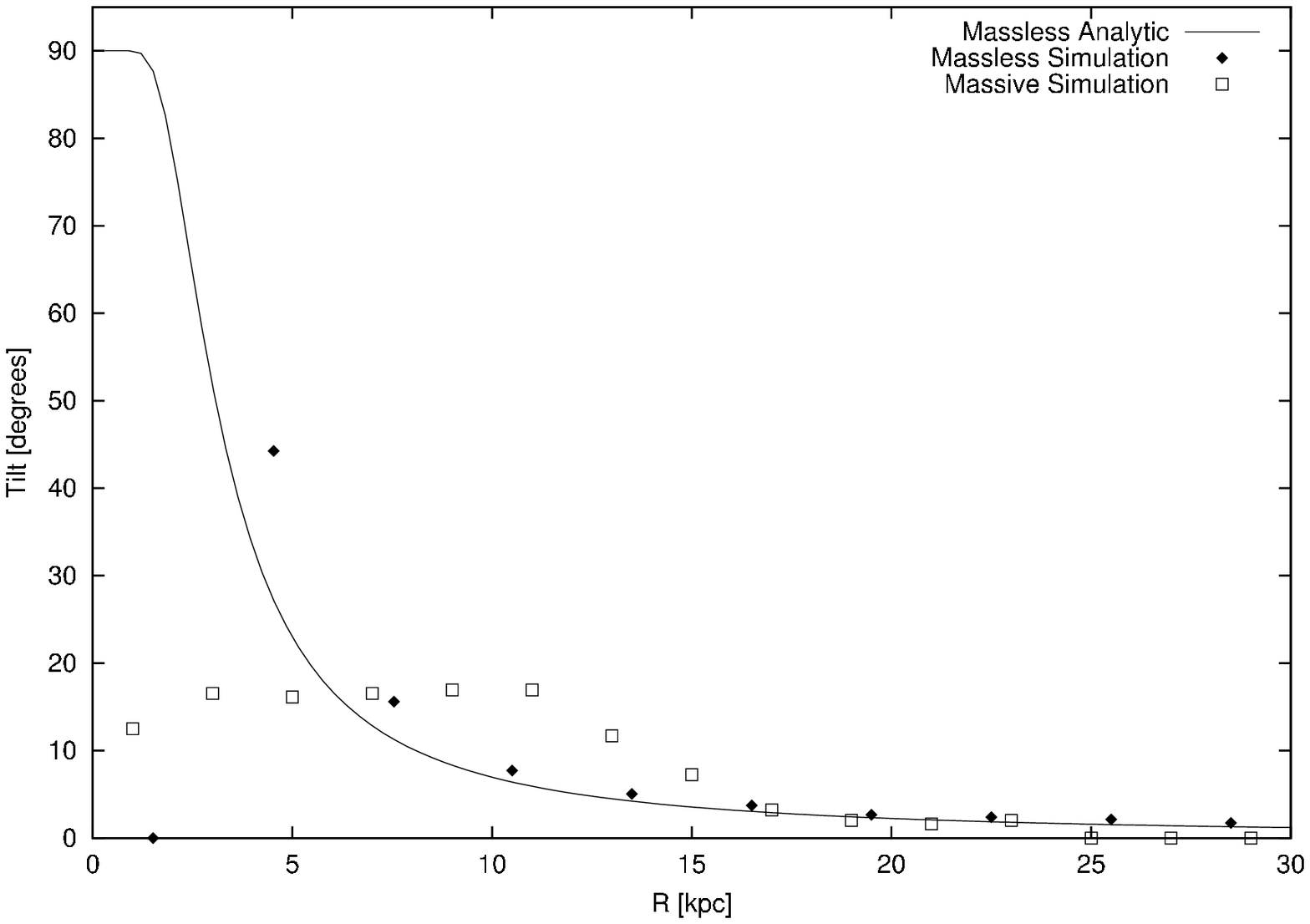}

\caption{Radial dependence of the tilt
  of the disk in Figure~\ref{simulation disk}, computed in spherical
shells 2~kpc wide. The solid line is the analytic 
prediction~(\ref{analytic better theta(r,t)})
and the solid diamonds are the massless N-body simulation,
as in Figure~\ref{plot massless theta(r,t)}.\label{plot tilt vs r selfgravity}}
\end{figure}

We binned the simulation particles into spherical shells of thickness
2~kpc and calculated
their tilts as in section~\ref{analytic vs simulations}. Figure~\ref{plot tilt vs r selfgravity}
shows a plot of the radial profile of the disk shown in Figure~\ref{simulation disk},
along with the massless simulation and analytic 
prediction~(\ref{analytic better theta(r,t)}) from
Figure~\ref{plot massless theta(r,t)},
which correspond to a massless disk in the same torque.
There are three important regimes: the central flat disk, the warp,
and the effectively massless outer region. The central 11~kpc is
all tilted \( 17\degr  \) from the original plane (we ignore the
innermost point because the torque is capped inside \( r_{0} \)),
and corresponds to the visually flat part of the disk. The effect
of the disk's self-gravity is to prevent the inner regions of the
disk from tilting as much as they would otherwise, while pulling the
outer regions of the disk to larger tilts. At the warp radius \( r_{w} \)=11~kpc,
the disk is no longer flat, and the disk warps back toward the original
plane. Finally, from 17~kpc to 23~kpc (the extent of the disk),
it follows the analytic prediction, acting like
a massless disk. This result agrees qualitatively with that of \citet{ostriker and binney89},
who found that regions of high surface density remain flat when torqued,
and the position of a warp is determined in part by a drop in surface
density.

\subsection{Warp evolution\label{warp evolution section}}

We now investigate how the disk evolves over time under the influence
of the torque. The general form of the warp is as above, but as the
disk evolves under the influence of the torque, the warp radius moves
out through the disk at a rate that depends on the mass of the disk.
At early times, typically before 200~Myr, the solid region is not
well developed and lies below the massless prediction.

\begin{figure}
\plotone{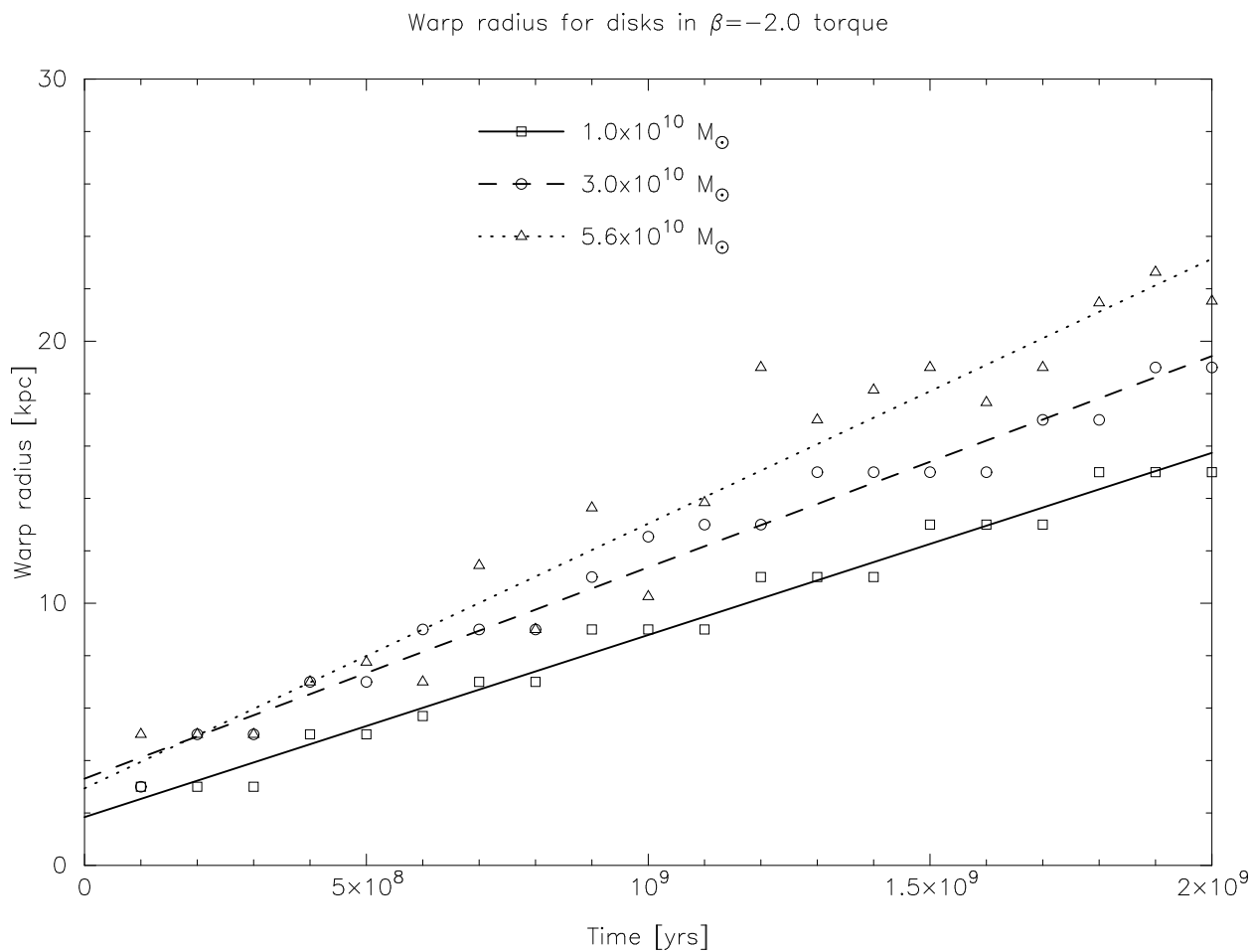}

\caption{Radial bins at which the disks exhibit their warps
every $10^8$~yr
  as the simulations evolve in time. Because the particles
were binned into spherical shells 2~kpc wide, the warp radii appear
quantized. The lines are linear least square fits.
This figure shows the simulations with \protect\( \beta =-2.0\protect \)
torques and disk masses \protect\( 1.0\times 10^{10}\: M_{\sun }\protect \),
\protect\( 3.0\times 10^{10}\: M_{\sun }\protect \), and \protect\( 5.6\times 10^{10}\: M_{\sun }\protect \)
represented by the squares, circles, and triangles respectively.\label{rw2.0 figure}}
\end{figure}

\begin{figure}
\plotone{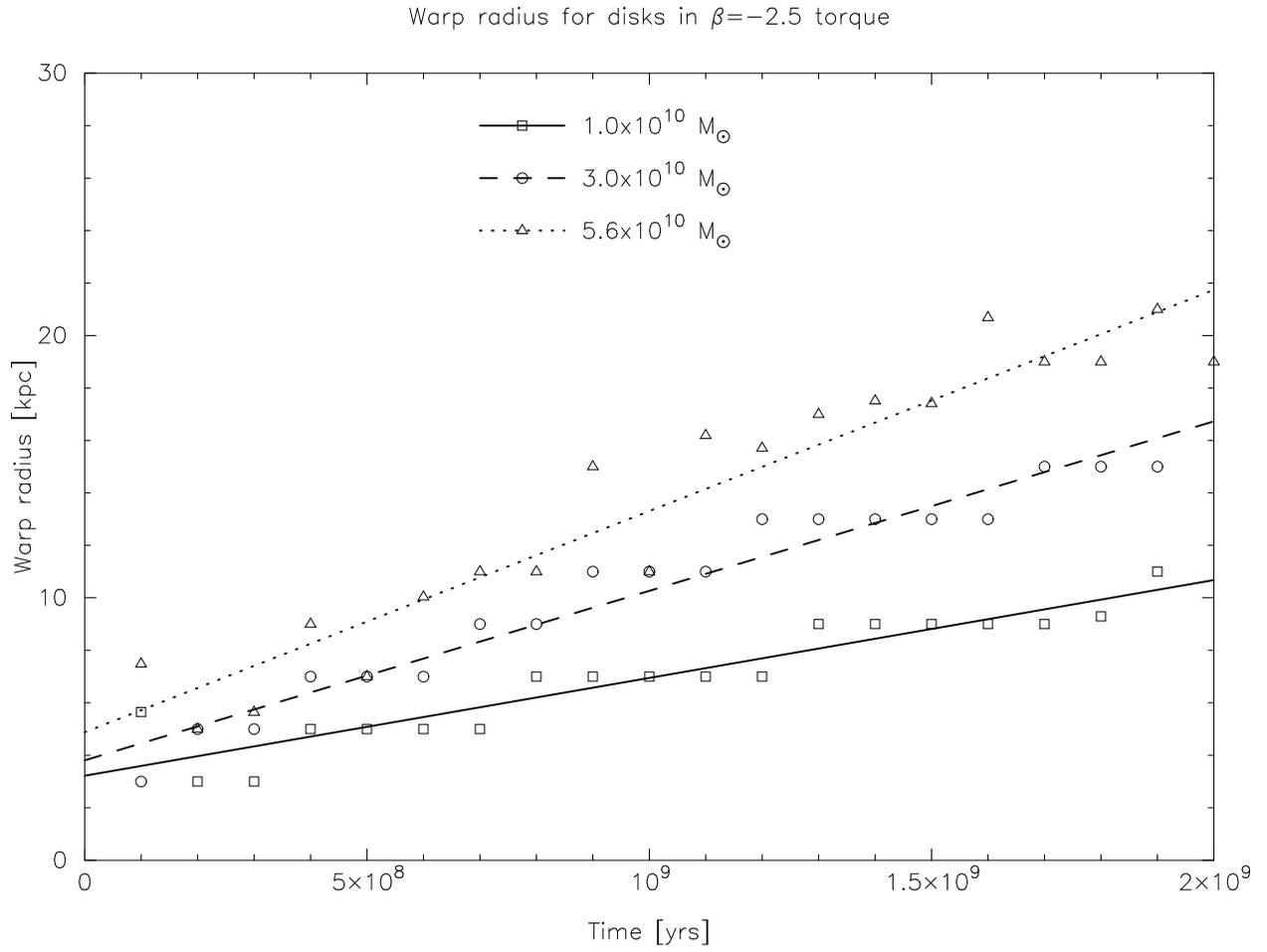}

\caption{As in Figure~\ref{rw2.0 figure} but for simulations with \protect\( \beta =-2.5\protect \)
torques.\label{rw2.5 figure}}
\end{figure}

\begin{figure}
\plotone{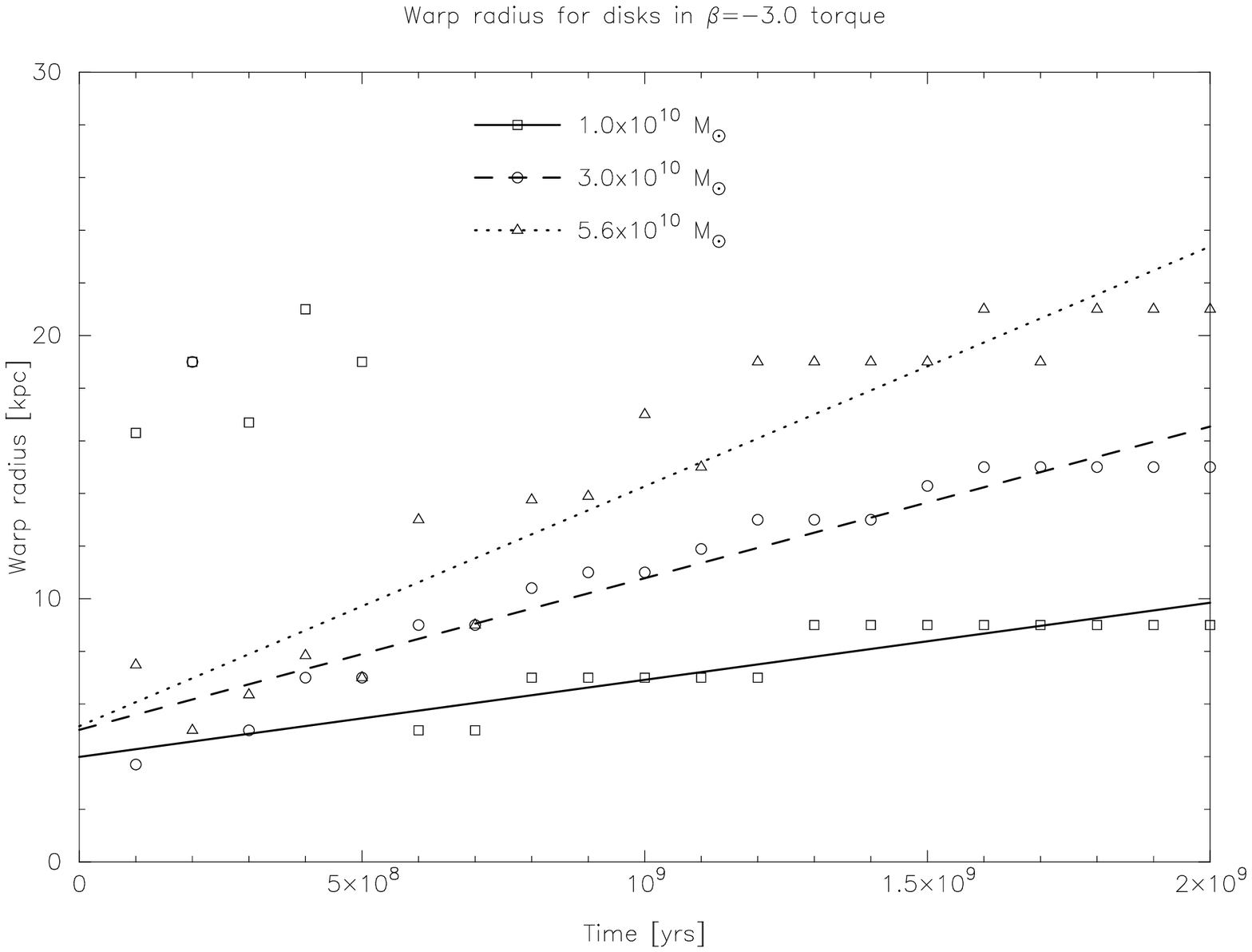}

\caption{As in Figure~\ref{rw2.0 figure} but for simulations with \protect\( \beta =-3.0\protect \)
torques. The warp takes time to develop, resulting in the noise apparent
early in the simulations, which were not included in the straight
line fits.\label{rw3.0 figure}}
\end{figure}

We developed an algorithm to automatically detect the warp radius.
The particle positions were stored every 40 timesteps.
For each of these outputs, the tilt angles were calculated in spherical
shells as in Figure~\ref{plot tilt vs r selfgravity}. We did a bootstrap
analysis to estimate the error in these angles: for each spherical
shell, we drew 10 random sets from the particles in that shell, calculated
the tilt of each bootstrap set, and then used the standard deviation
of those angles as an estimate of the error in the tilt of that bin.
The error depended on the number of particles in the shell, but was
typically less than 1\degr. 

Shells with errors greater than 0.5\degr\ were not analyzed, as these
generally had few particles and were dominated by numerical noise.
For each remaining shell with radius \( r \) and tilt \( \theta _{\mathrm{sim}} \),
we calculated the deviation \( \Delta \theta  \) between the simulation
tilt \( \theta _{\mathrm{sim}} \) and the analytic prediction \( \theta _{\mathrm{pred}}(r) \)
from 
equation~(\ref{analytic better theta(r,t)}). The average radius \( r \)
of the radial bin with the highest \( \Delta \theta  \) was then
considered to be the warp radius \( r_{w} \). This is the radius
at which the difference between the tilt of the massive disk and the
tilt of an equivalent massless disk is maximized. For example, in
Figure~\ref{plot tilt vs r selfgravity}, the highest \( \Delta \theta  \)
occurs at \( r=11\: \textrm{kpc} \) where the data point \( \theta _{\textrm{sim}}=17\degr  \)
and the predicted line \( \theta _{\textrm{pred}}=7\degr  \). This
corresponds exactly to the radial bin at the end of the plateau
in Figure~\ref{plot tilt vs r selfgravity}.
Examining a few randomly chosen outputs from each simulation showed
that in each case this automated \( r_{w} \) agreed with our intuition,
except at early times when there was no solid disk and the automated
\( r_{w} \) was dominated by noise.

Figures~\ref{rw2.0 figure}, \ref{rw2.5 figure},~and \ref{rw3.0 figure}
show the evolution of this warp radius for the set of simulations
with differing torque slopes $\beta$.
The squares, circles, and triangles correspond to simulations with disk
masses of 1, 3,~and \( 5.6\times 10^{10}\: M_{\sun } \) respectively.
The warp radii appear {}``quantized'' because of the 2~kpc-wide
radial bins used to calculate the tilt (see, e.g.~Figure~\ref{plot tilt vs r selfgravity}).
One point was taken every $10^8$ years, linearly interpolating $r_w$
from the neighbouring simulation outputs.
The early times before the warp has developed show up as noise
in the upper-left region of some of the figures,
particularly Figure~\ref{rw3.0 figure}. To quantify the growth of
the warp, straight lines were fit through the profiles.
While one might think that the lines ought to be constrained to
pass through the origin, since there is no warp at $t=0$, the data
do not support this
because the warp instability requires that the surface density of the disk
be below a certain critical surface density, as we demonstrate later.
We examined Figures~\ref{rw2.0 figure}--\ref{rw3.0 figure}
by eye and excluded the obviously noisy early times from the fits;
in practice, this meant excluding $t\le 5\times 10^8$ yr for the
$\beta=-3.0$, $M_d = 1.0\times 10^{10}\: M_{\sun}$ simulation and
$t\le 2\times 10^8$ yr for the $\beta=-3.0$, $M_d = 3.0\times 10^{10}\:
M_{\sun}$ simulation.
We ended the simulations after 2~Gyr, the time when
solid region reached the edge of the particle disk in some simulations,
at which point the algorithm to find \( r_{w} \) just detects the
edge of the disk. The fits are shown as the solid, dashed, and dotted
lines for the simulations with disk masses of 1, 3,~and \( 5.6\times 10^{10}\: M_{\sun } \)
respectively.

\begin{figure}
\plotone{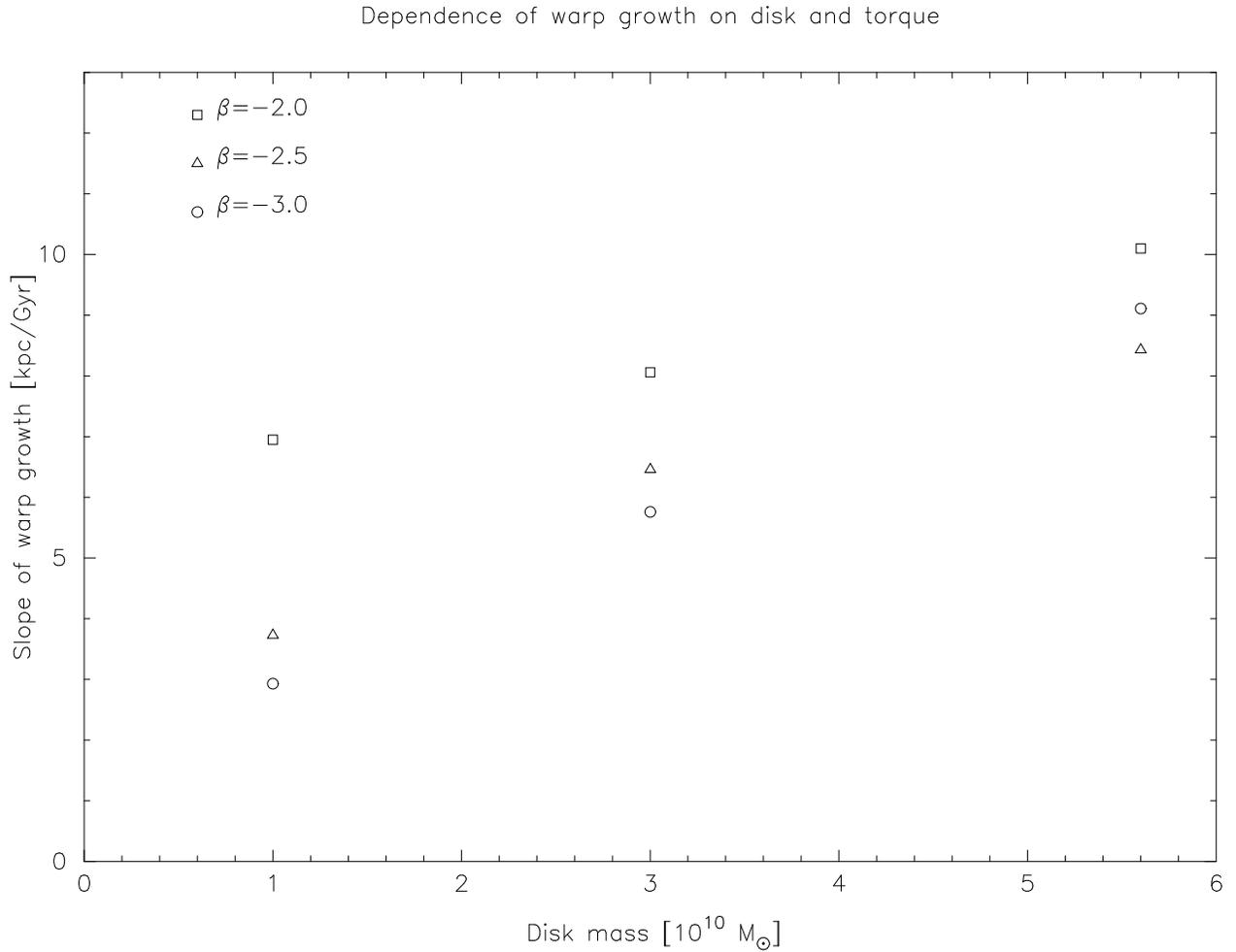}

\caption{Rate that the warp moves out through the disk (i.e. the slopes of
the linear fits in Figures~\ref{rw2.0 figure}--\ref{rw3.0 figure})
as a function of disk mass, with different torque slopes \protect\( \beta \protect \)
denoted by the different symbols. The effect of the slope on the rate
of warp growth is ambiguous, but there is a clear trend that more
massive disks have faster-growing warps.\label{warp growth}}
\end{figure}

The slope of the linear fit measures how quickly the warp grows. This
could depend both on the mass of the disk, which affects the local
surface density at the warp radius, and the slope of the torque, which
determines the strength of the torque at the warp radius. Figure~\ref{warp growth}
shows the rate that \( r_{w} \) grows for each simulation. There is an
apparent trend that the warps in higher mass disks move through
the disk faster than in
lower mass disks, but the scaling does not appear simple.
There is some indication
from Figure~\ref{warp growth} that the warps from shallower torque laws
(which have stronger torques at the warp radii) grow faster than those
with steep torques, but the situation is not clear.

\begin{figure}
\plotone{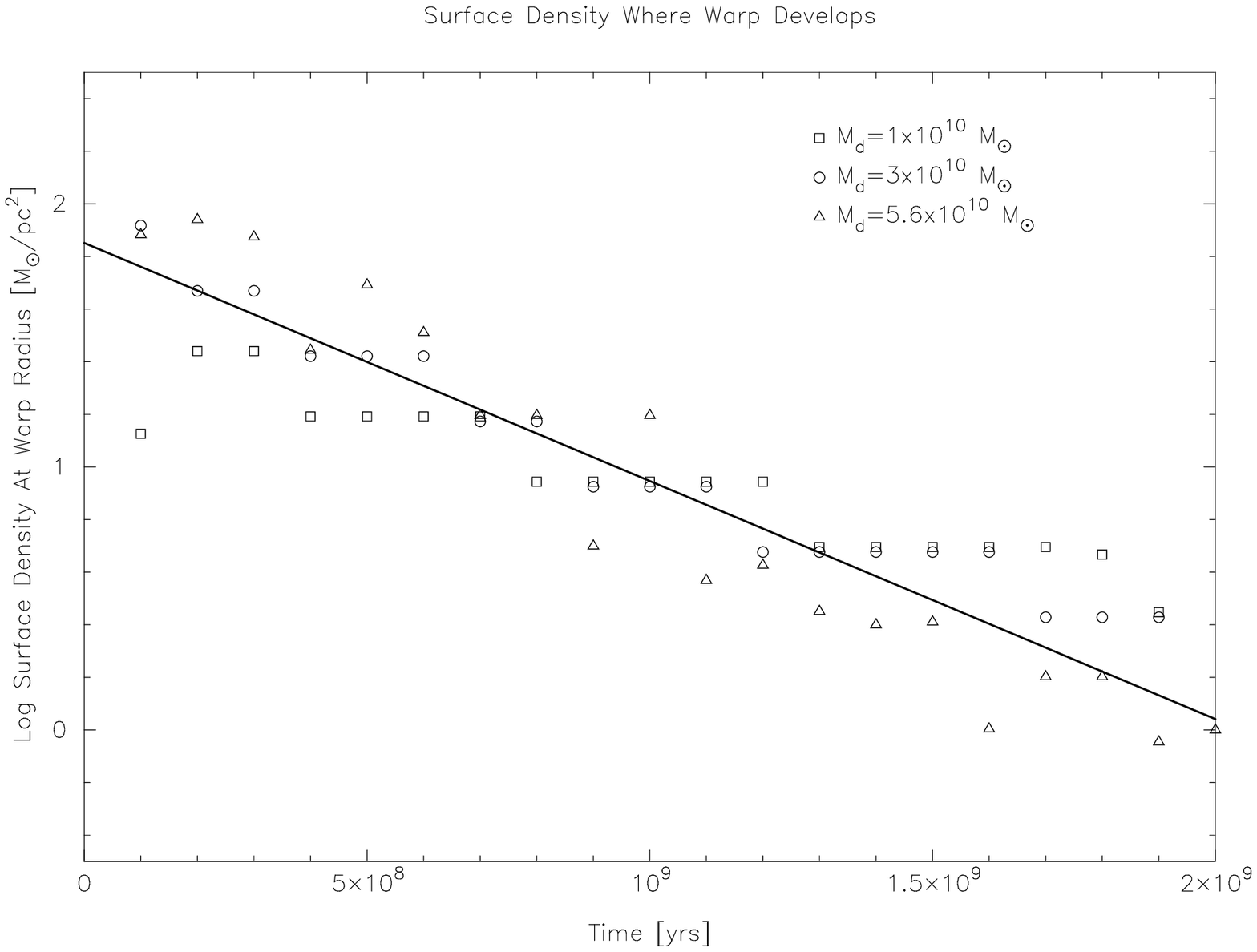}

\caption{Surface densities at the warp radii as the
simulations evolve. As the warp moves out through the disk, the local
surface density at that radius falls according to 
equation~(\ref{sigma(r)}).
The squares, circles, and triangles show simulations with disk masses
of 1,~3,~and~\protect\( 5.6\times 10^{10}\: M_{\sun }\protect \)
respectively in a \protect\( \beta =-2.5\protect \) torque.
The line is the fit (\ref{sigma warp(t)}).  Compare
this with Figure~\ref{rw2.5 figure}, which shows the radial evolution
of the warp for the same set of simulations.
The local surface density at the warp is similar for all
models at all times.\label{plot surface density}}
\end{figure}

\begin{figure}
\plotone{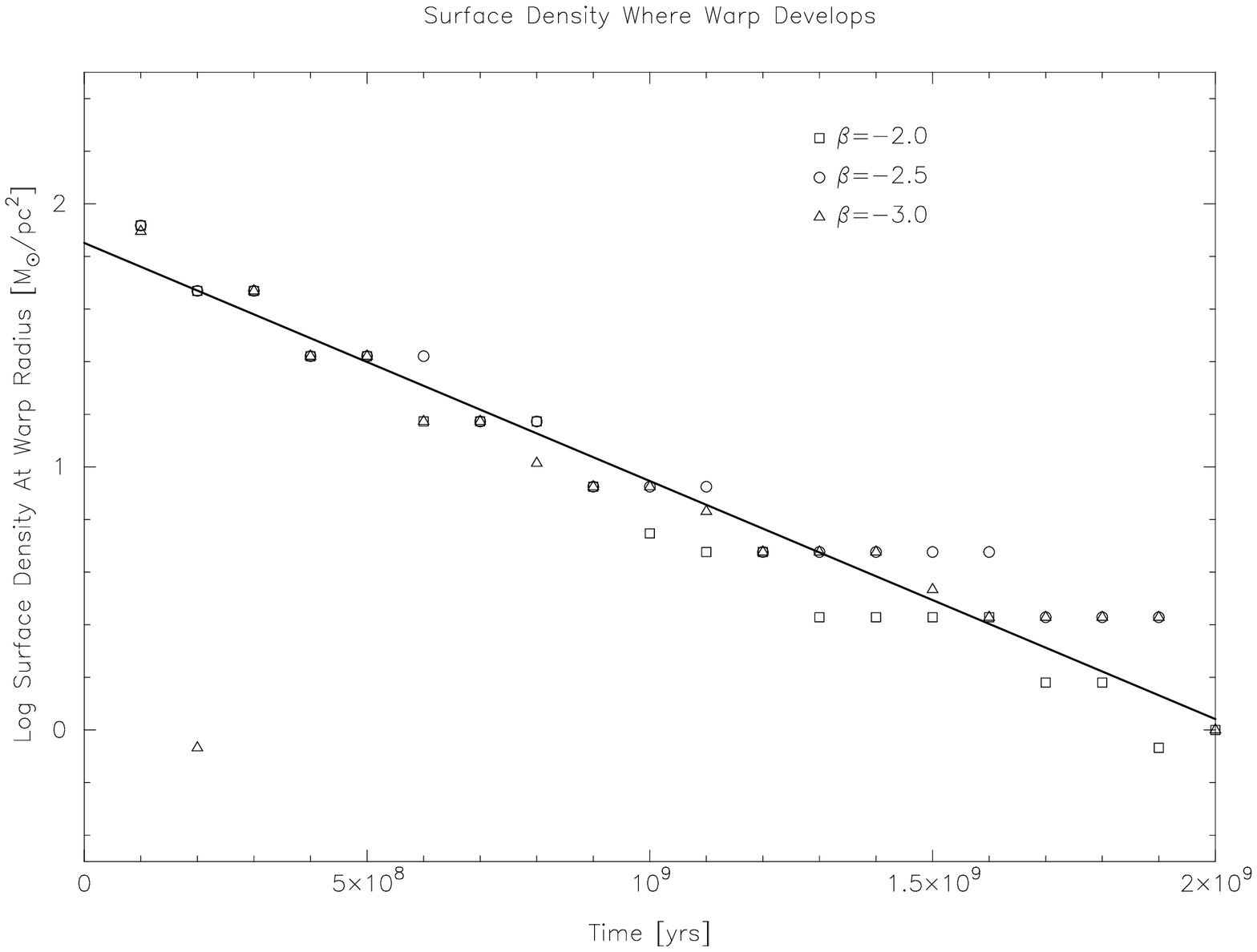}
\caption{As in Figure~\ref{plot surface density}, but plotting
$3.0\times 10^{10}\: M_{\sun}$ disk simulations
  with different torque slopes $\beta$.\label{plot sigma beta}}
\end{figure}

The effect of the disk mass on the rate of the warp growth implies
that the self-gravity of the disk is important to the formation of
the warp. In a massive disk, particles at different radii are coupled
to each other gravitationally and act to keep the disk locally flat.
This suggests that the crucial factor that determines where the warp
develops is the local surface density of the disk. \citet{ostriker and binney89}
examined the effect of a slewing disk potential on a set of self-gravitating
rings and found qualitatively that regions of high surface density react like a
solid body, but that warps can occur where the surface density is
lower. \citet{hofner and sparke94} noted that in most cases the group
speed of bending waves in a disk of surface density \( \Sigma (r) \)
and angular rotation velocity \( \omega (r) \) is\[
c_{g}=\frac{\pi G\Sigma (r)}{\omega (r)},\]
and therefore the time for a warp to settle at a given radius
would be inversely
proportional to the surface density at that radius.

To test this, we translate the warp radii of the simulations
into local surface densities using
\begin{equation}
\label{sigma(r)}
\Sigma (r_{w})=\frac{M_{d}}{4\pi r_{e}^{2}}e^{-r_{w}/r_{e}}
\end{equation}
for a disk of mass \( M_{d} \) and exponential scale length \( r_{e} \).
Figure~\ref{plot surface density} shows the surface density at the
warp radius as a function of time for the three disks of different
mass in a $\beta=-2.5$ torque (i.e. for the same simulations shown
in Figure~\ref{rw2.5 figure}), while 
Figure~\ref{plot sigma beta} shows
it for a $3.0\times 10^{10}\: M_{\sun}$ disk
in torques with varying $\beta$.
The surface density at the warp radius falls as the warp moves
out through the disk.
The local surface
densities at the warp \emph{at a given time} are quite similar for
all disk masses, and have even smaller scatter for different
torque laws.
It appears that the local surface density is the important
parameter for determining the warp radius at a given time.

The evolution of the warp is well described by a decaying exponential
\begin{equation}
\label{sigma warp(t)}
\Sigma(r_w,t) = \Sigma_{w_0} e^{-t/t_w}
\end{equation}
with $\Sigma_{w_0}=70\: M_{\sun}\: \mathrm{pc}^{-2}$ and
$t_w=480$ Myr, shown as the 
straight line in Figures~\ref{plot surface density}
and~\ref{plot sigma beta}.
For such a result to hold, the timescale $t_w$ should depend only on global
properties of the galaxy independent of disk mass or torque;
in particular, it can only depend on $\Sigma_{w_0}$,
the virial velocity of the galaxy 
\( v_{200}=175\, \mathrm{km}\, \mathrm{s}^{-1} \),
the exponential scale length of the disk $r_d=3.5\, \mathrm{kpc}$,
and/or the vertical scale height $h_z=325\, \mathrm{pc}$.
An interesting timescale that matches this is
the characteristic timescale for bending waves
\citep{hofner and sparke94}
one disk scale length away from $\Sigma_{w_0}$:
\begin{equation}
\label{tg equation}
t_g = \frac{r}{c_g} = \frac{v_{200}}{\pi G \Sigma_{w_0} e^{-1}}
\end{equation}
which is 490~Myr in these cases.

$\Sigma_{w_0}$ is the extrapolation of 
equation~(\ref{sigma warp(t)}) to
$t=0$, and is the critical surface density above which the disk
does not develop a warp.
At higher surface densities, the self-gravity of the disk is always
sufficient to keep the disk flat.
It is interesting that the extrapolation to $t=0$, when there
is no torque, gives a finite well-defined value for the warp
surface density. This is why the linear $r_w(t)$ fits do not go
through the origin and suggests that at this surface density the
disk is marginally unstable to warping.
We do not expect to see warps occurring at surface densities
higher than $70\: M_{\sun}\: \mathrm{pc}^{-2}$.
In the Milky Way, the warp begins at or
slightly beyond the solar circle \citep{binney92}, while recent
Hipparcos determinations of the Milky Way surface density in
the solar neighbourhood give $\Sigma_0 = 40\: M_{\sun}\: \mathrm{pc}^{-2}$
\citep{creze et al98}. It is encouraging that this is less
than $\Sigma_{w_0}$. In external galaxies, warps begin between
25~and 26.5~mag arcsec${}^{-2}$
\citep{briggs90}, which for a mass to light ratio
in $B$ of $1\: M_{\sun}/L_{\sun}$ gives surface densities
of between 1.8--7.0 $M_{\sun}\: \mathrm{pc}^{-2}$, ignoring
projection effects and extinction.
This is also consistent with a picture in which warps only
can occur at surface densities below $\Sigma_{w_0}$.

\section{Summary}

We studied how a typical galactic disk reacts to torques expected
from lying in misaligned dark matter halos. Our main findings are
the following:

\begin{itemize}
\item Cosmological N-body simulations suggest that galactic disks will be
misaligned with the mass distribution of the dark matter halo in which
they are embedded. Because of this misalignment, the halo exerts a
perpendicular gravitational torque on the disk of the form\[
\tau (r)=\tau _{0}\left( \frac{r}{1\, \mathrm{kpc}}\right) ^{\beta }\]
with typical values of \( \tau _{0}=10^{-30}\, \mathrm{s}^{-2} \)
and \( \beta =-2.5 \). This gravitational torque is strong enough
to have a significant effect on the entire disk within a Hubble time.
\item The timescale for an orbit to tilt in response to this torque rises
with radius, so inner portions of the disk will realign themselves
first, resulting in a warped disk. For a massless disk of stars in
circular orbits, a good estimate of how tilted the disk is at a radius
\( r \) after a time \( t \) is\[
\theta (r,t)=2\cos ^{-1}\left[ \frac{1}{\sqrt{2}}\frac{1+e^{t/t_{0}(r)}}{\sqrt{1+e^{2t/t_{0}(r)}}}\right] \]
where the timescale \( t_{0}(r) \) is defined as \[
t_{0}(r)=\frac{2v_{c}(r)}{\tau _{0}r_{0}}\left( \frac{r}{r_{0}}\right) ^{-\beta -1}\]
and is twice as long as in the case where the orbital radii are
not allowed to vary.

\item Massive disks depart from this due to the self-gravity of the inner
portions of the disk. The disk is kept flatter where the local surface
density is high than for a massless disk. The radius inside which
the disk is flat (which corresponds to the radius where the warp would
be considered to start if the disk were observed) grows with time.
The warp grows at a rate of between 3~and 10~kpc~Gyr\( ^{-1} \).
The rate depends on the mass of the disk, but is relatively insensitive
to the torque parameters.

\item More massive disks have faster-growing warps,
since only at large radius is the surface density sufficiently
low that its self-gravity cannot maintain the flat disk.
The surface density at which the warp occurs is well-described by
\[
\Sigma_w(t) = \Sigma_{w_0} e^{-t/t_w}
\]
where $\Sigma_{w_0}=70\: M_{\sun}\: \mathrm{pc}^{-2}$ and
$t_w=480$~Myr, independent of disk mass and torque slope.
$\Sigma_{w_0}$ is the maximum surface density at which a warp can 
occur, and marks the point where a disk is marginally unstable
to warping. $t_w$ appears to coincide with the timescale of
bending waves one disk scale length away from $\Sigma_{w_0}$:
\[
  t_g = \frac{v_{200}}{\pi G \Sigma_{w_0} e^{-1}}.
\]

\end{itemize}
In the future, we plan to analyze high resolution cosmological simulations
to determine the torques due to realistic dark matter halos. This
will enable us to distinguish the torque due to the misaligned disk,
that due to substructure in the halo, and that due to the external
tidal field by modeling individual halos. A further direction that
we have not yet explored is sampling cosmological simulations at regular
intervals up to the present day to replace the static torque force
used in this work with a torquing field that changes over time in
a cosmologically-motivated manner.

\acknowledgements{
Several figures were produced using the WIP package \citep{morgan95}.
This work was supported by NSF grants 9807151 and PHY99-0749, and
NSERC grant PGSB-233028-2000. MS is an Alfred~P.~Sloan Fellow and a
David~and Lucile~Packard Fellow.}


\begin{thebibliography}{Reshetnikov \& Combes(1998)}
\bibitem[Athanassoula et al(1987)]{athanassoula et al 87}Athanassoula, E., Bosma, A., \& Papaioannou, S. 1987, \aap, 179, 23
\bibitem[Battaner et al(1990)]{battaner et al}Battaner, E., Florido, E., \& Sanchez-Saavedra, M.L. 1990, \aap, 236,
1
\bibitem[Binney(1992)]{binney92}Binney, J.J. 1992, \araa, 30, 51
\bibitem[Binney \& Tremaine(1987)]{BT}Binney, J.J., \& Tremaine, S.D. 1987, Galactic Dynamics (Princeton:
Princeton University Press)
\bibitem[Bosma(1981)]{bosma81}Bosma, A. 1981, \aj, 86, 1791
\bibitem[Briggs(1990)]{briggs90}Briggs, F. 1990, \apj, 352, 15
\bibitem[Broeils \& Courteau(1997)]{broeils and courteau97}Broeils, A.H., \& Courteau, S. 1997, in ASP Conf. Ser. 117, Dark and
Visible Matter in Galaxies and Cosmological Implications, ed. M. Persic,
Salucci, P. 74
\bibitem[Christodoulou et al(1993)]{christodoulou et al93}Christodoulou, D.M., Tohline, J.E., \& Steiman-Cameron, T.Y. 1993,
\apj, 416, 74
\bibitem[Cole \& Lacey(1996)]{cole and lacey96}Cole, S., \& Lacey, C. 1996, \mnras, 281, 716
\bibitem[Courteau \& Rix(1999)]{courteau and rix99}Courteau, S., \& Rix, H.-W. 1999, \apj, 513, 561
\bibitem[Cr\'ez\'e et al(1998)]{creze et al98}Cr\'ez\'e, M., Chereul, E., Bienaym\'e, O., \& Pichon, C. 1998, \aap, 329, 920
\bibitem[de Blok \& McGaugh(1997)]{de blok and mcgaugh97}de Blok, W.J.G., \& McGaugh, S.S. 1997, \mnras, 290, 533
\bibitem[Debattista \& Sellwood(1999)]{debattista and sellwood99}Debattista, V.P., \& Sellwood, J.A. 1999, \apjl, 513, L107
\bibitem[Dekel \& Shlosman(1983)]{dekel and shlosman83}Dekel. A., \& Shlosman, I. 1983, in IAU Symp. 100, Internal Kinematics
and Dynamics of Galaxies, ed. E. Athanassoula (Dordrecht: Kluwer),
187
\bibitem[Diplas \& Savage(1991)]{diplas and savage91}Diplas, A., \& Savage, B.D. 1991, \apj, 377, 126
\bibitem[Dubinski \& Carlberg(1991)]{dubinski and carlberg91}Dubinski, J., \& Carlberg, R.G. 1991, \apj, 378, 496
\bibitem[Frenk et al(1985)]{fwed85}Frenk, C.S., White, S.D.M., Efstathiou, G., Davis, M. 1985, \nat, 317, 595
\bibitem[Garc\'ia-Ruiz et al(2000)]{garcia-ruiz et al00}Garc\'ia-Ruiz, I., Kuijken, K., Dubinski, J. 2000, preprint (astro-ph/0002057)
\bibitem[Hernquist(1993)]{hernquist93}Hernquist, L. 1993, \apjs, 86, 389
\bibitem[Hofner \& Sparke(1994)]{hofner and sparke94}Hofner, P., \& Sparke, L.S. 1994, \apj, 428, 466
\bibitem[Katz(1991)]{katz91}Katz, N. 1991, \apj, 386, 325
\bibitem[L\'opez-Corredoira et al(2002a)]{lopez-corredoira et al02a}L\'opez-Corredoira, M., Betancort-Rijo, J., \& Beckman, J.E. 2002a, \aap, 386, 169
\bibitem[L\'opez-Corredoira et al(2002b)]{lopez-corredoira et al02b}L\'opez-Corredoira, M., Cabrera-Lavers, A., Garz\'on, F., Hammersley, P.L. 2002b, \aap, in press (astro-ph/0208236)
\bibitem[Lynden-Bell(1965)]{lynden-bell65}Lynden-Bell, D. 1965, \mnras, 129, 299
\bibitem[Morgan(1995)]{morgan95}Morgan, J.A. 1995, WIP - An Interactive Graphics Software Package, in Astronomical Data Analysis Software and Systems IV, ed. R.A. Shaw, H.E. Payne, and J.J.E. Hayes. PASP Conference Series 77, 129
\bibitem[Navarro et al(1996)]{NFW96}Navarro, J.F., Frenk, C.S., \& White, S.D.M. 1996, \apj, 462, 563
\bibitem[Navarro et al(1997)]{NFW97}Navarro, J.F., Frenk, C.S., \& White, S.D.M. 1997, \apj, 490, 493
\bibitem[Okumura et al(1993)]{grape3}Okumura, S.K., Makino, J., Ebisuzaki, T., Fukushige, T., Ito, T.,
Sugimoto, D., Hashimoto, E., Tomida, K., \& Miyakawa, N. 1993, \pasj,
45, 329
\bibitem[Ostriker \& Binney(1989)]{ostriker and binney89}Ostriker, E.C., \& Binney, J.J. 1989, \mnras, 237, 785
\bibitem[Porciani et al(2001)]{porciani et al02}Porciani, C., Dekel, A., \& Hoffman, Y. 2002, \mnras, 332, 339
\bibitem[Quinn \& Binney(1992)]{quinn and binney92}Quinn, T., \& Binney, J. 1992, \mnras, 255, 729
\bibitem[Reed(1996)]{reed96}Reed, B.C. 1996, \aj, 111, 804
\bibitem[Reshetnikov \& Combes(1998)]{reshetnikov and combes98}Reshetnikov, V., \& Combes, F. 1998, \aap, 337, 9
\bibitem[Ryden(1988)]{ryden88}Ryden, B.S. 1988, \apj, 329, 589
\bibitem[Sackett(1997)]{sackett97}Sackett, P. 1997, \apj, 483, 103
\bibitem[Sackett(1999)]{sackett99}Sackett, P. 1999, in ASP Conf. Ser. 182, Galaxy Dynamics, ed. D. Merritt,
J. A. Sellwood, \& M. Valluri, 393
\bibitem[Sparke \& Casertano(1988)]{sparke and casertano88}Sparke, L.S., \& Casertano, S. 1988, \mnras, 234, 873
\bibitem[Steinmetz(1996)]{steinmetz96}Steinmetz, M. 1996, \mnras, 278, 1005
\bibitem[Toomre(1983)]{toomre83}Toomre, A. 1983, in IAU Symp. 100, Internal Kinematics and Dynamics
of Galaxies, ed. E. Athanassoula (Dordrecht: Kluwer), 177
\bibitem[Tsuchiya(2002)]{tsuchiya02}Tsuchiya, T. 2002, New Astronomy, 7, 293
\bibitem[Tully \& Fisher(1977)]{tully fisher}Tully, R.B., \& Fisher, J.R. 1977, \aap, 54, 661
\bibitem[van den Bosch et al(2002)]{vdb et al02}van den Bosch, F.C., Abel, T., Croft, R.A.C., Hernquist, L., \& White, S.D.M. 2002, \apj, in press (astro-ph/0201095)
\bibitem[Warren et al(1992)]{warren et al92}Warren, M.S., Quinn, P.J., Salmon, J.K., \& Zurek, W.H. 1992, \apj, 330, 519
\bibitem[Weinberg(1998)]{weinberg98}Weinberg, M.D. 1998, \mnras, 299, 499
\end{thebibliography}
\end{document}